\documentclass[12pt]{iopart}
\usepackage{color}
\usepackage{iopams}
\usepackage{mathrsfs}
\usepackage{latexsym}
\usepackage{subfigure, graphicx}
\usepackage{caption}
\usepackage{float}
\usepackage{hyperref}
\usepackage{ulem}
\usepackage{tabularx}
\usepackage{academicons}
\usepackage{xcolor}
\usepackage{hyperref}
\usepackage{svg}
\usepackage[T1]{fontenc}
\usepackage[utf8]{inputenc}

\begin{document}

\def\prd{{\it Phys. Rev.} D~}
\def\PRL{{\it Phys.Rev.} Lett~}
\def\apjl{{\it Astrophys. J.} Lett~}
\def\apj{{\it Astrophys. J.}}
\def\PRD{{\it Phys. Rev.} D~}
\def\CQG{{\it Class. Quantum Grav.}}
\def\aaps{{\it A\&AS~}}
\def\pasj{{\it PASJ~}}
\def\mnras{{\it MNRAS~}} 
\def\aapr{{\it A\&ARv~}}
\def\aap{{\it A\&A~}}
\def\araa{{\it A\&RAA~}}
\def\na{{\it New Astronomy~}}
\def\ptp{{\it Progress of Theoretical Physics~}}

\newcommand\hlc[2]{\bgroup\markoverwith
  {\textcolor{#1}{\makebox[0.75pt]{\rule[-.5ex]{1pt}{2.5ex}}}}\ULon{#2}}

\definecolor{Periwinkle}{RGB}{204, 204, 255}
\newcommand{\rh}[1]{\textsf{\hlc{Periwinkle}{ RH: #1}}}
\newcommand{\gda}[1]{\textsf{\hlc{green}{ GDA: #1}}}

\newcommand{\orcid}[1]{\href{https://orcid.org/#1}{\textcolor[HTML]{A6CE39}{\aiOrcid}}}

\hyphenation{in-fras-truc-ture}

\title{DataVault: A Data Storage Infrastructure for the Einstein Toolkit}

\author{Yufeng Luo$^{1,4}$, Roland Haas$^1$, Qian Zhang$^2$, Gabrielle Allen$^{1,3,4,5}$}
  
\address{$^1$ National Center for Supercomputing Applications, University of Illinois at Urbana-Champaign, Urbana, Illinois, 61801, USA}
\address{$^2$ David R. Cheriton School of Computer Science, University of Waterloo, Waterloo, ON, Canada  N2L 3G1}
\address{$^3$ College of Education, University of Illinois at Urbana-Champaign, Urbana, Illinois, 61801, USA}
\address{$^4$ Department of Astronomy, University of Illinois at Urbana-Champaign, Urbana, Illinois, 61801, USA}
\address{$^5$ Department of Mathematics and Statistics, University of Wyoming, Laramie, Wyoming, 82071, USA}


\begin{abstract}
Data sharing is essential in the numerical simulations research. We introduce a data repository, DataVault, that is designed for data sharing, search and analysis. A comparative study of existing repositories is performed to analyze features that are critical to a data repository. We describe the architecture, workflow, and deployment of DataVault, and provide three use-case scenarios for different communities to facilitate the use and application of DataVault. Potential features are proposed and we outline the future development for these features.
\end{abstract}
%
%
%
%
%

\clearpage


\section{Introduction} \label{sec:introduction}

Numerical computing is a fundamental part of modern scientific research, with computer codes and their resulting data sets becoming ever more complex and costly to produce, maintain and support. Collaborations between different research groups around code development and use are now commonplace. These collaborations can be closed or open depending on whether new members can easily join and contribute to data sets and computer codes. Whichever form the collaboration takes, a platform to easily store, share and analyze data generated by or consumed by the collaboration is critical to the success of these collaborations.

Outside the field of astrophysics, data repositories are already  widely used
in different disciplines. For example, the Protein Data Bank
(PDB)~\cite{pdb:web} is a repository of information about the 3D structures of
large biological molecules, which enables users to deposit, search for,
download, visualize and analyze PDB data. Zenodo~\cite{zenodo:web}, as another
examples is a
generic data repository that supports a broad wide range of domains
and disciplines.

This paper provides a case study of a new data platform, ``DataVault''~\cite{datavault:web,datavaultcode:web}, built to support the Einstein Toolkit community~\cite{EinsteinToolkit:2020_11}. The Einstein Toolkit is a community-driven open-source software platform of core computational tools to advance and support research in relativistic astrophysics and gravitational physics. The user base includes more than 300 scientists distributed across 200 research groups worldwide who use different software components of the  Einstein Toolkit to model scenarios in relativistic astrophysics, including black holes, neutron stars, supernovae and gravitational waves.

DataVault is designed around a number of motivating use cases that are also common in other fields of computational science:
\begin{itemize}
\item Fostering collaboration across a geographically dispersed and loosely organized group of collaborators with a semi-private repository,
where any collaborator can upload data sets. Users are able to query to see if data is already available for a particular parameter choice and then retrieve and analyze the results.
\item Sharing the data produced by a scientific (sub-)community through a centrally hosted data repository to a wider science community. As an example, the initial
target group for DataVault is the interaction between the numerical relativity (NR)
community and waveform modelling groups that relies on sharing of data sets
produced in NR simulation which are used to calibrate
semi-analytic waveform
models.
\item Providing data sets produced within the scientific community to a wider,
non-scientist audience via a well-defined, easy-to-use interface. This enables
citizen science efforts as showcased, for examples, in the Citizen
Science~\cite{citizenscience:web} and Galaxy Zoo~\cite{galaxyzoo:web}
projects
that make data accessible to a larger user base,
such as high-school students and students at non-PhD granting universities.
\end{itemize}

In order to address these usage cases,  we have designed DataVault, a web-based, domain specific 
data storage facility that can be easily used and deployed by groups interested in
sharing their data products. We are specifically targeting the NR
waveform community which exemplifies all three usage cases outlined above.

For the Einstein Toolkit community, we design the DataVault repository to achieve multiple functionalities for different researchers. We provide a large set of waveforms spanning a so far unexplored region of parameter space for scientists involved in gravitational waveform (GW) modeling. Users have a full set of functions to help them better share their data and metadata associated with it. As an open-source collaboration, DataVault and Einstein Toolkit project team members include faculty, research staff, as well as students world wide. This international collaboration shapes DataVault to better match current research needs and be more adaptive to new research directions.

Key functionality in DataVault to facilitate efficient data sharing among scientists
includes:
\begin{itemize}
\item Advanced metadata searching using knowledge of the type of
stored metadata  in DataVault allows for quick selection of data sets that users
want to study further. Moreover, DataVault, as an open-source project, can
be modified by users to match their needs, potentially contributing the functionality
back to the public DataVault code. 
\item Public data accessibility is an important factor
underpinning results in a publication and enabling further research based on prior work. DataVault enables users to publish data sets they
used in publications, so any readers can download the data to verify and further
analyze their results conveniently. Data citation and validation of data sets are
important factors in promoting data sharing. To this end DataVault provides a unique
identifier for each data set as well as for collections of multiple data sets.
DataVault supports both closed installations where data sets can only be uploaded
by a pre-approved set of collaborators as well as open installations that allow 
users to register and upload their own data.
\item An easy to use, well-defined interface provides the non-scientific public with access to scientific data sets for outreach purposes.
Data stored in DataVault is made discoverable and interested users without in-depth
familiarity with the research groups can easily access the data. 
\end{itemize}

\paragraph{Organization of this paper.}
This manuscript is organized as follows: section 2 reviews existing data
repositories used in numerical astrophysics, section 3 introduces DataVault's
framework, section 4 outlines future directions for DataVault and section 5
contains our main conclusions and summary.

\section{Existing repositories}\label{sec:existing_repositories}

This section provides an non-exhaustive study, summarized in Table~\ref{table:repocompare}, of existing data platforms that have been built by individual numerical research groups or collaborations~\cite{sxscatalog:web, gtcatalog:web, corecatalog:web, ritcatalog:web}. In addition the study includes generic platforms, such as Zenodo~\cite{zenodo:web} and the Illinois Data Bank~\cite{illinoisdatabank:web}.  The study aimed to identify the capabilities and functionalities available in 8 different repositories, comparing them using the following criteria.

\begin{description}
    \item[Open source] Is the source code of the repository infrastructure available under an Open Source license? An open source licensed repository encourages customization based on specific needs, and facilitates the sustainability of the data repository when its maintenance passes to new developers.
    \item[Domain specific] Is the repository a generic data repository or domain-specific repository that targets a specific user community?
    \item[Persistent Identifier (PID)] Does the repository provide a persistent identifier for each data set that can be used as a citation? Data will be retrieved long after it was uploaded initially, making a persistent identifier mandatory.
    \item[Metadata extraction] When a user uploads the data set, can the repository process the data set with custom operations, such as metadata extraction and grouping multiple data sets into a collection?
    \item[Numerical search] Does the repository provide numerical search functionality beyond basic textual search? Users should be able to search for numerical ranges in metadata.
    \item[Downloadable metadata] Is the full set of searchable metadata available for download and offline processing? Complex queries beyond the provided search facilities can be implemented by experienced users using the raw metadata to select only specific data sets for download.
    \item[Add collaborators] Does the repository support multiple people to enter metadata and upload files for a submission? In this way, research groups can invite collaborators to contribute to a data set.
\end{description}

\subsection{Repositories in the numerical astrophysics community}

Multiple groups active in the field of numerical astrophysics have made their
simulation results available to the public. In particular for the numerical
relativity community that is the initial target community of DataVault, the
the Numerical Injection Analysis (NINJA)~\cite{Aylott:2009ya} and Numerical
Relativity
Analytical Relativity (NRAR)~\cite{NRAR:2013} projects brought
together numerical relativists and gravitational waveform modellers. They, for
the first time, defined a common data format used by multiple groups and all
existing repositories of NR waveforms use formats descended
from those defined by NINJA and NRAR.

This section provides a short review of the existing repositories of
gravitational waveforms. All currently existing repositories are maintained by
individual research groups or collaborations, employing relatively
straightforward web-frontends to the data hosted. They typically provide
functionality to find waveforms based on parameters describing the simulation,
as well as ways to download the full set of metadata describing the data in
the repository. 

For each repository we provide a short overview and summarize our findings in
table~\ref{table:repocompare}. Since these are private repositories using, at
least for some functionality, proprietary, non-disclosed code, we cannot
report on the mechanism available to user to populate the database.

\begin{table}[htbp]
\small%
\begin{tabular}{|*9{c|}}
\hline
 Repository &
 DataVault &
 GT &
 RIT &
 SXS &
 CoRE &
 Zenodo &
 IDB &
 FRDR \\
 \hline
 Open source &
 Yes &
 No &
 No &
 Yes\textsuperscript{1} &
 No &
 Yes &
 No &
 No \\
 \multicolumn{1}{|p{2.3cm}|}{\centering{}Domain specific} &
 Yes &
 Yes &
 Yes &
 Yes &
 Yes &
 No &
 No &
 No \\
 \multicolumn{1}{|p{2.3cm}|}{\centering{}Download metadata} &
 Yes &
 Yes\textsuperscript{2} &
 Yes\textsuperscript{3} &
 Yes\textsuperscript{4} &
 No &
 No &
 No &
 No \\
 PID &
 Custom &
 Custom &
 Custom &
 DOI &
 Custom &
 DOI &
 DOI &
 DOI \\
 \multicolumn{1}{|p{2.3cm}|}{\centering{}Metadata extraction} &
 Yes &
 N/A &
 N/A &
 N/A &
 N/A &
 No &
 No &
 No \\
 \multicolumn{1}{|p{2.3cm}|}{\centering{}Numerical search} &
 Yes &
 No &
 Yes &
 Yes &
 No &
 No &
 No &
 No \\
\multicolumn{1}{|p{2.3cm}|}{\centering{}Add collaborators} &
 Yes &
 N/A &
 N/A &
 N/A &
 N/A &
 No &
 No &
 Yes \\
\hline
\end{tabular}
\caption{Comparison of existing repositories.
\textsuperscript{1} uses Zenodo code,
\textsuperscript{2} single text file,
\textsuperscript{3} tar archive,
\textsuperscript{4} single json file.
}
\label{table:repocompare}
\end{table}

\paragraph{Georgia Tech Catalog of Gravitational Waveforms}
The Georgia Tech (GT) Catalog of Gravitational Waveforms~\cite{gtcatalog:web,Jani:2016wkt}
is a closed, domain-specific data catalog currently containing 452 distinct
waveforms from
binary black holes (BBH) simulations formerly maintained by the
NR group at Georgia Tech.  The catalog is organized as a
set of
individual simulations in a 12-dimensional parameter space, including initial parameters
of binary black holes and the resulting remnant black hole, along with a unique identifier
and an internal name.  Keyword search and sorting facilities are provided by
proprietary JavaScript code on the website.  The data for a given simulation
is available in individual Numerical Relativity Injection
(NRI)~\cite{Schmidt:2017btt} files stored on GitHub.  Example scripts to
read these files are provided via GitHub~\cite{gtscripts:web}.

\paragraph{Rochester Institute of Technology binary black hole simulations catalog}

The Rochester Institute of Technology (RIT) numerical relativity group's closed
catalog
contains 777 binary black holes waveforms~\cite{ritcatalog:web,
Healy:2019jyf}.  The catalog
is organized as a set of individual simulations in a 20-dimensional parameter
space,
including initial parameters of binary black holes and remnant black hole
along with a unique identifier and simulation parameters. Keyword and
numerical range based search and sorting facilities are provided by
proprietary JavaScript code on the website. The data for one or multiple
simulation(s) can be downloaded.  Simulation data is available as NRI format
files stored at RIT. In addition metadata is provided in NRAR format files~\cite{NRAR:2013} files, a
predecessor of the NRI format.

\paragraph{SXS Gravitational Waveform Database at Cornell}

The closed Simulating eXtreme Spacetimes (SXS) Gravitational Waveform
Database~\cite{sxscatalog:web, Boyle_2019} consists of 2018 black hole and
neutron star merger simulations.
The catalog if organized as a set of individual simulations in a
90-dimensional parameter
space, including initial parameters of binary black holes or neutron stars and
remnant black hole or neutron star along with a unique identifier and
simulation parameters. Keyword and numerical range based search and sorting
facilities are provided by proprietary JavaScript code on the website. The
data for individual simulations can be downloaded.  Metadata is provided in
custom ASCII files based on the NRAR format.  Simulation data is available as
custom HDF5 format files stored in a Zenodo instance hosted by Caltech
library.  Example scripts to read these
files are provided via GitHub~\cite{sxsscripts:web}.

\paragraph{Computational Relativity (CoRe)}

The Computational Relativity (CoRe) catalog~\cite{corecatalog:web,
Dietrich:2018phi} of
waveforms consists of 367 binary neutron star merger simulations.  The closed
catalog
is organized as a set of individual simulations in a 12-dimensional parameter space,
including initial parameters of binary neutron stars and remnant black hole or
neutron star along with a unique identifier and simulation parameters. Keyword
based search and sorting facilities are provided by proprietary JavaScript
code on the website. The data for individual simulations can be downloaded.
Metadata is provided in custom ASCII files defined on the catalog website.
Simulation data is available as custom HDF5 format files stored in a GitLab
instance hosted at one of the CoRe collaborators' institution.

\subsection{Generic repositories}

Recognition of data as valid research products and the need to provide
ways to share data and track academic credit for data products has lead to
the development of data sharing solutions on both the institutional level and 
spanning institutions. Data retention policies adopted by funding
agencies~\cite{nsfdatasharing:web} have furthered this development. In this section, one international, one institutional, and one federated (national) platform is introduced, respectively.

\paragraph{Zenodo}

Zenodo~\cite{zenodo:web} is a general purpose
open source~\cite{zenodorepo:web} research publishing infrastructure and
repository developed under the European OpenAIRE program and operated by
CERN. It allows researchers to upload, and share all forms of
research outputs such as data sets, research software, reports, and any other
research related digital artifacts. Each data set is owned by a single user
and fine grained access control is not provided. Upon publishing, a Digital Object
Identifier (DOI) is minted and assigned to the scholarly record. Keyword based
search functionality is provided by Zenodo and individual submissions or
individual files in submissions can be downloaded.

\paragraph{Illinois Data Bank}

The Illinois Data Bank (IDB)~\cite{illinoisdatabank:web} is a proprietary
institutional data repository which provides access to Illinois
research data. IDB allows individual researchers to upload, and share all
forms of
research outputs such as data sets, research software, reports, and any other
research related digital artifacts but does not allow to restrict access to a
subset of users. Upon publishing, a DOI
is minted and assigned to the scholarly record. Keyword based
search functionality is provided by IDB and individual submissions or
individual files in submissions can be downloaded.
Uploads are limited to researchers at the University of Illinois, while
downloads are available to all researchers.

\paragraph{Federated Research Data Repository}
Federated Research Data Repository (FRDR)~\cite{frdr:web} is a proprietary
platform for Canadian researchers to deposit and share digital research data
and to facilitate discovery. Being a federated design FRDR supports fine
grained access control to data sets. Upon publishing, a
DOI is minted and assigned to the scholarly record. FDDR provides keyword
based search functionality.  Uploads are limited to Canadian researchers,
while downloads are availalbe to all researchers.

\subsection{DataVault}
By combining advantages from domain-specific and generic data platforms, we design, develop and deploy a domain-specific data repository for sharing and archiving NR simulation data, featuring collaboration, open data and open science and meanwhile promoting FAIR~\cite{FAIR:web} (Findability, Accessibility, Interoperability, and Reuse) data principles. The following section discusses how features of DataVault ensure it provides each of the desired features mentioned in table~\ref{table:repocompare}.

\section{DataVault description}\label{sec:datavault}

DataVault is built by and initially for the Einstein Toolkit community. Its design  is based on the   community's needs as shown by its key features in table~\ref{table:repocompare}. This section gives a high-level overview of DataVault's structure and functionalities, as well as how the functionalities are applied to different use-case scenarios.

\subsection{Architecture and workflow}

DataVault is implemented as a web-based application connecting to a storage and
search engine. Its user interface provides easy management and
discovery of data sets. This section describes architecture and design features in detail.

Data access, processing and storage are among the most critical factors when developing a research data repository \cite{reporeqmatrix}. When users upload data sets, DataVault processes them through multiple pipelines, which include metadata extraction, and collection assignment. Structure and workflow of the DataVault is shown in the figure~\ref{fig:structure}.
\begin{figure}[htbp]
    \centering
    \includegraphics[scale=0.8]{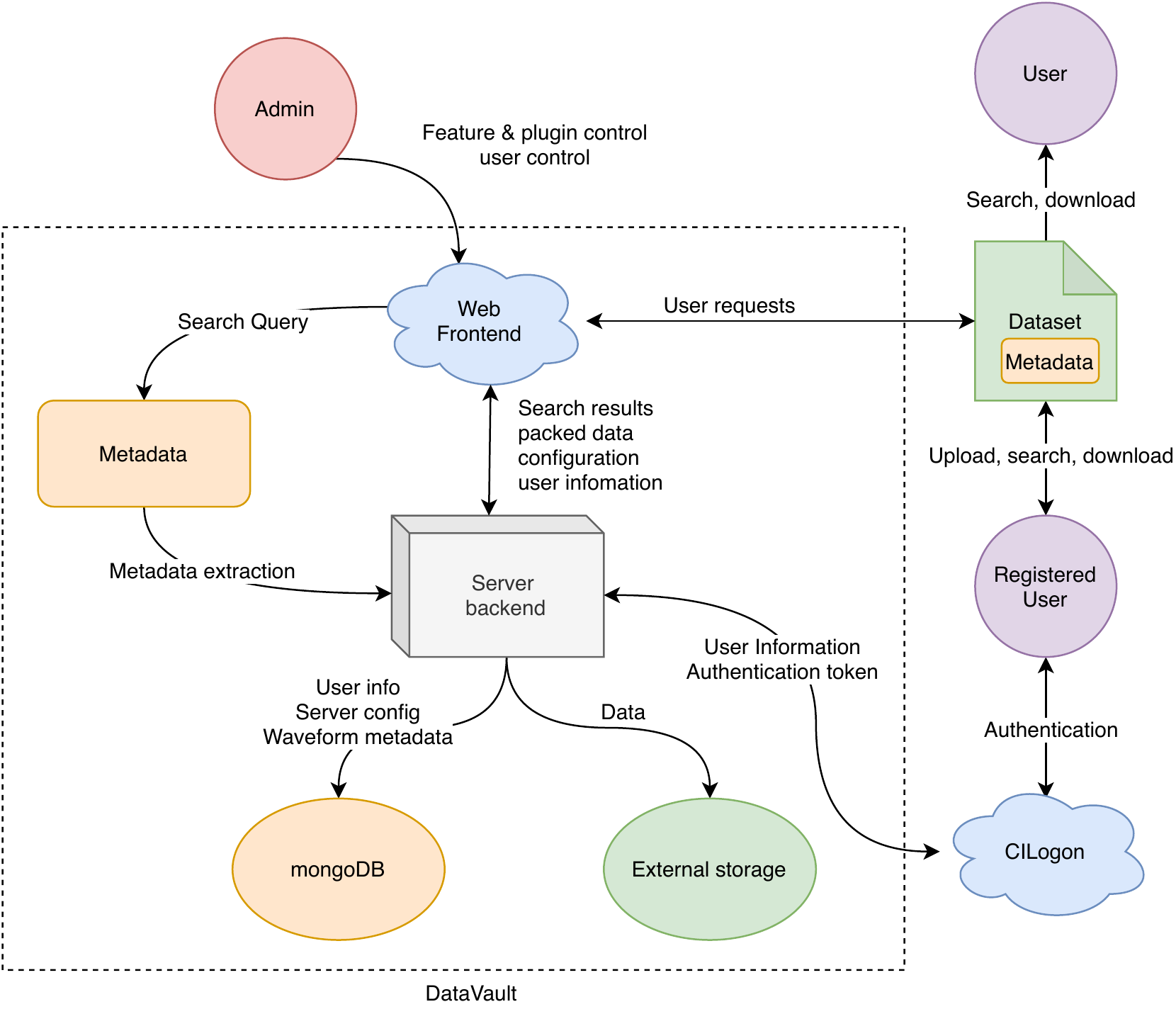}
    \caption{DataVault structure and workflow. DataVault components are enclosed within the black box. All interactions are indicated by arrows. Users are authenticated using CILogon, and only then can they uploaddata sets from DataVault. When a user uploads the waveform, the metadata is extracted on the server side and stored in MongoDB. All data and metadata is stored in external storage to preserve the data in case the Docker image needs to be rebuild. Users can send search queries either using the web user interface or a command line interface. The server parses the search query, retrieves and returns the requested information to user.}
    \label{fig:structure}
\end{figure}

To advance open science and ensure wider distribution, DataVault is developed based on the open-source data analytics platform \textit{Girder}~\cite{girder:web}. Girder includes basic functionalities, such as data set upload and download, collection creation, user group assignment and registration control.
Girder can be divided into two components: the core and plugins. The core component of Girder implements the fundamental framework for both the user interface front-end and server back-end. It constructs the critical API endpoints to let additional plugins provide features and acquire information, via event triggers such as ``upload'' and ``new user registration''. All extended functionality is implemented as Girder plugins. In general, a plugin is composed of front-end and back-end parts. The front-end is implemented using JavaScript and serves as an interface between the user and the back-end. The back-end is implemented using Python, it handles requests sent by the front-end and makes API calls to retrieve the information from the core component and process them.

Girder is chosen because it has a complete set of the functionalities required to effectively build the data repository with desired features. Girder is designed with security and extensibility in mind via the separation into core and plugin components. Therefore, research groups using DataVault can implement new features as additional plugins to Girder. Furthermore, Girder provides detailed documentation and has an active user community for discussion and support, and is actively developed and maintained.

DataVault is implemented as a set of Girder plugins. These plugins include metadata extraction, advanced metadata search, and CILogon as an extension to the existing OAuth2 login plugin.

\paragraph{Metadata extraction.} Metadata describes simulation data sets. The metadata fields in the table~\ref{tab:nri_format} are critical to waveform data sets. They specify physical initial conditions for the simulation, such as masses, initial separation and initial orbital velocity which are key parameters of interest to waveform modellers and numerical relativists.
\begin{table}[htbp]
    \centering
    \begin{tabular}{|c|p{6cm}|c|c|}
        \hline
          attribute name & \centering physical parameters & type & range \\
          \hline
          mass1 ($M_1$) & \centering mass of more massive object & float & [0, $\infty$)\\
          mass2 ($M_1$) & \centering mass of less massive object &float &  [0, $\infty$)\\
          eta ($\eta$) & \centering the symmetric mass ratio &float & (0, 1]\\ 
          spin1 x,y,z ($\chi_{1, (x,y,z)}$) & \centering dimensionless spin vector of object 1 in NR frame & float &  \\
          spin2 x,y,z ($\chi_{2, (x,y,z)}$) & \centering dimensionless spin vector of object 2 in NR frame & float & \\
         LNhat x,y,z ($\hat{L}_{n, (x,y,z)}$) & \centering Newtonian orbital angular momentum unit vector & float & \\
         nhat x,y,z ($\hat{n}_{(x,y,z)}$)& \centering the orbital separation unit vector & float & [-1, 1] \\
         Omega ($M\Omega$) & \centering dimensionless orbital frequency & float & (0, $\infty$) \\
         eccentricity ($e$) & \centering estimated eccentricity & float & [0, 1] \\ 
         mean\_anomaly & \centering estimated mean anomaly & float & -1 for N/A, [0, 1] \\
         \hline 
    \end{tabular}
    \caption{Numerical relativity injection~\cite{Schmidt:2017btt} format metadata fields understood by DataVault. Spin1, spin2, LNhat, nhat are 3D vector quantities and have a separate metadata field for each x, y, z component. }
    \label{tab:nri_format}
\end{table}
The set of metadata also contains a user chosen identifier for the waveform. DataVault provides extensive support for metadata presentation and extraction. When a user uploads waveforms, DataVault automatically extracts metadata from the uploaded files and verifies the format of the metadata to ensure it adheres to NR Injection format~\cite{Schmidt:2017btt} specifications. 
As a domain-specific data repository with known metadata, DataVault offers advanced semantic search functionality, which is introduced in the next paragraph.

\paragraph{Search plugin.}
To help users find waveforms more efficiently, DataVault provides an advanced search plugin. Users search for waveforms using the metadata fields shown in table~\ref{tab:nri_format}. When searching, users specify a numerical range for the metadata attributes they are interested in. If a lower limit or upper limit not given, then the search assumes it to be unrestricted.  Users can immediately see search results and download either all or a subset of the waveforms from within the result display. Figure~\ref{fig:search_UI} shows the search user interface and results display.
\begin{figure}
    \centering
    \includegraphics[scale=0.4]{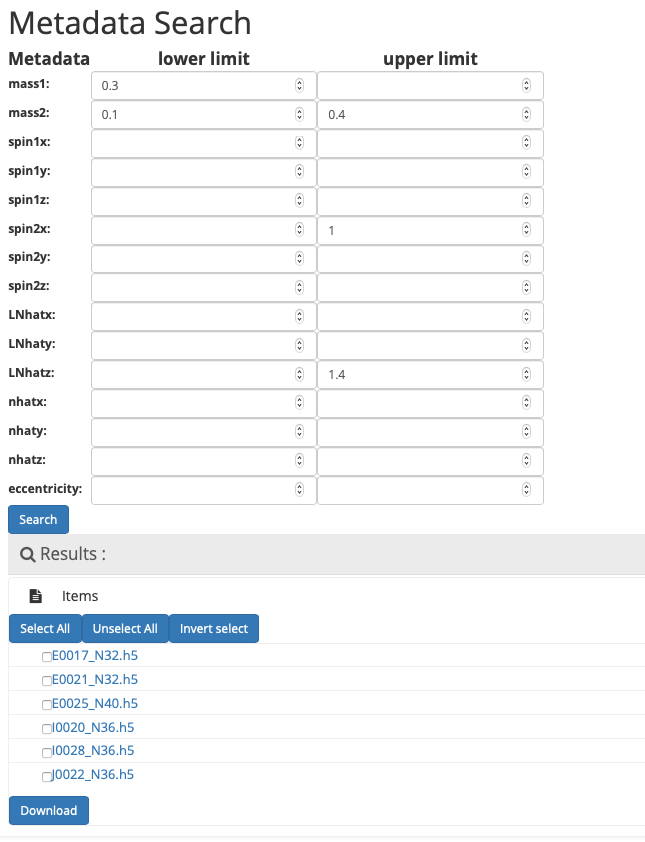}
    \caption{Search plugin user interface. Users input lower and upper limit in the search box. The results section shows all waveforms with metadata values within the specified numerical range. Users then select individual ones or all waveform files for download.}
    \label{fig:search_UI}
\end{figure}


\paragraph{CILogon based authentication.}
To ensure only authorized users can upload data and access private data,  authentication plugins require community users to log. 
DataVault primarily relies on CILogon for
institutional logins. CILogon is developed for academic users, it allows users to use their institutional credentials to log in without need ingto create a separate DataVault account. Thus DataVault benefits from trust in the identities reported by CILogon. As an example this allows a research group to easily grant access to new members without having to perform their own identity verification. Generic Oauth2 login, such as GitHub and BitBucket, is also
supported for external collaborators whose home institutions are not participating in CILogon.
An additional layer of access control enables that administrators
approve access of registered users before new users can upload data.

\paragraph{Data set collections.}
DataVault is a platform for sharing data, including published data and private data for ongoing research. To facilitate these goals, DataVault supports creating data collections consisting of multiple data sets. Access to all data sets in a collection is controlled at the collection level which provides a convenient way to organize related data sets. For example research groups that share data among members, can create data collections and grant access to their group members only but none other. In this way, data stored online remains private to the research group. Collections also organize data sets into categories to simplify searching for and downloading of data sets matching common criteria.

\subsection{Deployment}
\begin{figure}[htbp]
    \centering
    \includegraphics[scale=0.8]{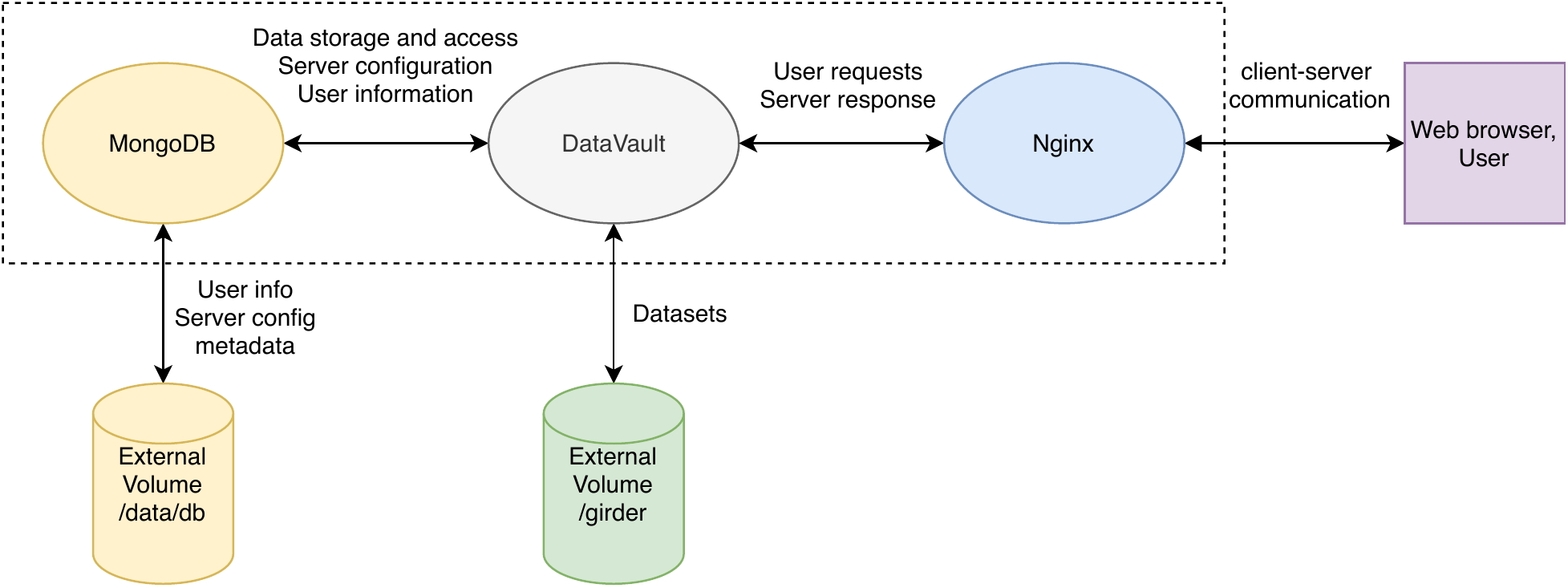}
    \caption{DataVault container setup. Each ellipse represents a container. Users communicate with the Nginx container which forwards requests to the DataVault container. MongoDB also resides in a separate container. All data and configuration files are stored outside of the containers.}
    \label{fig:docker}
\end{figure}
DataVault is containerized to achieve portability and speed up deployment. Containerization is achieved using Docker~\cite{docker:web}, which is open-source, lightweight and portable. Both the Dockerfile used to build the DataVault image and the docker-compose file to build the full server infrastructure consisting of DataVault, MongoDB~\cite{mongodb:web} and Nginx~\cite{nginx:web} are available for download~\cite{datavaultcode:web}. Figure~\ref{fig:docker} shows the interaction between the components.

The MongoDB volume is mounted externally for data persistence since data stored in the container is destroyed when the container stops running. Thus, in case the server operation is interrupted and the running docker container is shut down, the external mounting volume can preserve the server configurations, users information and the data set stored in the DataVault.

\subsection{DataVault use-cases and application}
DataVault's features described above target multiple scenarios, three examples are provided in this section to illustrate some use-cases for the application of DataVault into different research tasks.

\paragraph{For numerical relativity and waveform modeling communities}
DataVault directly benefits groups using the Einstein Toolkit, both as producers of waveform data sets and consumers of these data sets. Einstein Toolkit users can store the data using the DataVault instance hosted at NCSA. These users store waveforms generated for published papers, as using the unique PIDs generated for data set provenance and citation. A sample set of 89 waveforms generated by the NCSA gravity group serves as a demonstration for how data sets can be arranged in collections. Ownership of the data sets is shared among all NCSA gravity group members with access controlled by DataVault. Non-NCSA groups interested in this data sets access it using read-only permission. DataVault's advanced search functionality immediately provides a convenient
means for groups modeling gravitational waves using semi-analytical methods
to select and download numerical waveform data sets for calibration and
testing of their models.

\paragraph{For numerical astrophysics community}
A DataVault instance is hosted at NCSA for public use and other research groups can choose to self-host a DataVault instance to store data locally. They use the collections to group data sets of
interest and share access to the collections among group members.
On the other hand, numerical groups whose data sets are not supported by public
DataVault instances, or who desire a ``branded'' instance of DataVault,
make use of the open-source nature of DataVault
to access the source code and modify
it to accommodate their needs. In particular, they adapt metadata extraction and advanced search
functionality, if their data sets have a clear structure and metadata for
searching.

\paragraph{For public and education use}
Data stored in NCSA's DataVault instance is open for public us. For students
and instructors at institutions without a numerical astrophysics research
program, or even for high-school educators, DataVault provides a way to access
research data produced by supercomputing simulation. An instructor uses
DataVault to select data for their students to work, restricting
access to the collection to their respective class. Student upload simulation
results and share them among themselves.

\section{Sustainability}

DataVault is maintained by the Einstein Toolkit team at NCSA and there is a path for the platform to be supported by the Einstein Toolkit community as DataVault gains users. This section discusses the future plan for DataVault and how it will be supported in different phases of the operation as part of the Einstein Toolkit.

DataVault's future development is a iterative process consisting of
two alternating phases: development phases and maintenance phases. Each of
these phases
defines a set of objectives for DataVault. Although there is no clear boundary
between each phase, goals and objectives for each phase are individually
specified below.
\begin{description}
    \item[Development phase]
    In this phase, we focus on developing new functionality and 
    features for DataVault. This includes developing new plugins, A/B
    tests and gathering user feedback for further improvement and
    optimization.
    \item[Maintaining phase]
    The main objective of this phase is to keep the production DataVault in
    normal operation, and provide users with access to data. The team will not
    be developing some new features during this phase, and instead  gather
    feedback from users and discuss for further possibilities.
\end{description}

\subsection{Extensibility}
 DataVault builds on Girder, meaning that additional features can be easily implemented by modifying the plugins, as discussed in section \ref{sec:datavault}. DataVault's source code repository~\cite{datavaultcode:web} contains full developer's documentation of the existing plugins, considerably simplifying the task of adding new Girder plugins. DataVault being licensed under an open source license, the existing plugins serve as a starting point.

\subsection{Future work}
In the future we plan to add new features and optimization of current features based on community feedback. We plan to add more detailed online visualization, such as waveform plots and metadata visualization. 
DataVault's search functionality is key to its use and we will extend it to allow for user defined queries and more complex combinations of search terms. This will be an extension to the current advanced search functionality.

\section{Conclusion}\label{sec:conclusion}	
In this work, we presented a domain-specific data repository, DataVault, to facilitate collaborations by sharing data among relativity research groups and a wider non-scientists audience. To inform the fundamental framework design and collect necessary features for a potential data repository in NR community, a comparative review of 8 existing data repositories was conducted. 5 of the repositories are domain-specific and the other 3 are generic. We summarized their capabilities and compared them using a set of criteria that are necessary for research work in the numerical astrophysics community. 

In the comparative study we found that there was currently no single repository with all the features useful to the numerical astrophysics community.
Using the information collected in the study, we designed and developed a data repository, DataVault, that provides the identified features. An overview of DataVault’s architecture, deployment and extensibility was provided.
To illustrate the usefulness of DataVault, we investigated three use cases and discussed how DataVault can be applied in each scenario.
We conclude that DataVault enables an efficient, multi-purpose data sharing process for data intensive projects. 

\section{Acknowledgements}\label{sec:acknowledgements}
We thank Kacper Kowalik for helpful discussions when designing DataVault.
This research is part of the Blue Waters sustained-petascale computing
project, which is supported by the National Science Foundation (awards
OCI-0725070 and ACI-1238993) the State of Illinois, and as of December, 2019,
the National Geospatial-Intelligence Agency. Blue Waters is a joint effort of
the University of Illinois at Urbana-Champaign and its National Center for
Supercomputing Applications.
The eccentric numerical
relativity simulations used in this article were generated with the open
source, community software, the Einstein Toolkit on the Blue Waters
supercomputer and XSEDE (TG-PHY160053). This work was partially supported by
the NSF awards OAC-1550514 and OAC-2004879. We thank Ian Hinder
and Barry Wardell for the \texttt{SimulationTools}~\cite{SimulationTools:Web}
analysis package.

\appendix

\section{Waveform catalog}
To illustrate use of DataVault and speed up its adoption by the Einstein
Toolkit community we added the catalog of 89 eccentric NR
simulations reported in~\cite{Huerta:2019oxn} supplemented by an eccentricity
measurement using the method presented in~\cite{Habib:2019cui}. Waveforms were
postprocessed using \texttt{SimulationTools}~\cite{SimulationTools:Web}, and
converted to NRI file format~\cite{Schmidt:2017btt}.

This waveform catalog consists of non-spinning binary black hole merger
waveforms with mass ratio $1 \le q \le 10$ and eccentricities
$0 \le e_0 \le 0.18$ at fifteen waveform cycles before merger and contains
multipolar modes up to $\ell = 4$. All simulations used the Einstein
Toolkit~\cite{EinsteinToolkit:2020_11} using parameter files based
on~\cite{wardell_barry_2016_155394}.

DataVault makes all metadata stored in the HDF5 files accessible and
searchable through its user interface and allows subsets of the catalog
to be downloaded.

The catalog data set serves as a real world test case for DataVault's
usability as a waveform repository for the numerical astrophysics community.
\begin{itemize}
    \item Waveforms were produced by multiple members of the NCSA gravity
    group, making use DataVault user and group management functionality when
    controlling access to the data set while it was being prepeared.
    \item The large number of files require automatic extraction of metadata,
    describing the numerical simulations.
    \item The multi-dimensional parameter space spanned by mass ration and
    eccentricity makes it possible to define non-trivial subsets for example
    by selecting all circular waveforms or a sequence of increasingly
    eccentric waveforms for a fixed mass ratio.
    \item These waveforms are of interest to the larger gravitational waveform
    modelling community, with potential users outside of the research group at
    NCSA, thus demonstrating the usefulness of community repository.
\end{itemize}

\section{Eccentricity estimation and extraction}
The eccentricity of the NR simulations waveforms were extracted using the algorithm specified in~\cite{Habib:2019cui}. The algorithm found the eccentricity by passing the initial guess of the eccentricity through a coarse search and then followed by a fine search. We modified the code by adding a Newtonian initial guess for the eccentricity, based on the initial parameters from the metadata file created in each simulation. These parameters include the initial positions ($\vec{r}_1$, $\vec{r}_2$), initial linear momentum ($\vec{p}_1$, $\vec{p}_2$), and initial ADM mass ($m_1$, $m_2$). By using classical orbital mechanics, the Newtonian initial guess for eccentricity is:
\begin{equation}
    e_0 = \sqrt{1+\frac{2E_{sp}h_{sp}^2}{G M}}
\end{equation}
where $E_{sp}$ and $h_{sp}$ are the specific energy and specific angular momentum respectively, defined as 
\begin{eqnarray}
    E_{sp} &=& \frac{1}{\mu}\left(\frac{\|\vec{p}_1\|^2}{2m_1} + \frac{\|\vec{p}_2\|^2}{2m_2} - \frac{Gm_1 m_2}{\|\vec{r}_1 - \vec{r}_2\|}\right) \\
    h_{sp} &=& \frac{\|\vec{h}_1 + \vec{h}_2\|}{\mu}
\end{eqnarray}
$\vec{h}_1$ and $\vec{h}_2$ are the angular momentum of the two objects. These eccentricities were added to the metadata of the waveforms and presented online on the DataVault.

\section*{References}
\bibliographystyle{iopart-num}
\bibliography{references}

\end{document}